\documentstyle[12pt]{article}

\begin{document}
\begin{center}
{ \Large
Critical behavior of an absorbing phase transition \\
in an interacting monomer-dimer model
}

\vspace{ 5 mm}
Hyunggyu Park and Heungwon Park\\
Department of Physics\\
Inha University\\
Inchon 402-751, KOREA
\end{center}
\begin{abstract}
We study a monomer-dimer model with repulsive interactions
between the same species in one dimension. With infinitely strong
interactions the model exhibits a continuous transition from a
reactive phase to an inactive phase with two equivalent absorbing
states. Static and dynamic Monte Carlo simulations show that the
critical behavior at the transition is different from the
conventional directed percolation universality class but is
consistent with that of the models with the mass conservation
of modulo 2. The values of static and dynamic critical exponents
are compared with those of other models.
We also show that the directed percolation
universality class is recovered when a symmetry-breaking
field is introduced.
\end{abstract}
\vspace{3 mm}
\begin{center}
\end{center}

A monomer-dimer(MD) model was introduced by Ziff, Gulari, and Barshad
to describe the oxidation of carbon monoxide on catalytic surface
\cite{ZGB}.
In this model, a monomer ($CO$) adsorbs onto a single vacant site, while
a dimer ($O_2$) adsorbs onto a pair of adjacent vacant sites and then
immediately
dissociates. A nearest neighbor of adsorbates, comprised of a dissociated
$O$ atom and an $CO$ atom, reacts and forms a $CO_2$ molecule and desorbs
from the metal surface.
In two dimensions, as the $CO$ gas pressure is lowered, the system
undergoes a first-order
transition from a $CO$-saturated inactive phase into a reactive
steady state and then a continuous transition into a $O_2$-saturated
inactive phase. This continuous transition
is shown to be
in the same universality class as the directed percolation (DP)
\cite{Cardy,Jan,Grass82}. In one dimension, there is only a first-order
phase transition between two saturated phases.

Motivated by the monomer-dimer model, many related lattice models have
been formed to study nonequilibrium phase transitions, e.g.~the
contact process(CP) \cite{Harris},
the A model \cite{Dick88}, the interacting monomer-monomer model
(IMM) \cite{Park,Zhuo931},
and so forth. The CP and A model are single-component models
(there are two choices at a given site: vacant/occupied
or healthy/diseased), while
the MD and IMM model are multi-component models (three choices:
vacant/monomer/dimer or the other monomer).
A common feature of these models,
whether single-component or multi-component,
is that they exhibit a phase transition from a reactive
phase into an inactive phase of a {\em single} absorbing state.
The resulting
critical behaviors are classified into the category of the so-called
``DP conjecture'' \cite{Grass82,Grin} which depicts that models
exhibiting a continuous transition to a single absorbing state
generally belong to the universality class of the directed
percolation.

The universality class for models with a single
absorbing state is well established.  But few studies have
been made for models with more than one absorbing states.
Recently
Jensen and Dickman \cite{Jen932,Jen933} have
extensively studied some nonequilibrium
lattice models with infinitely many absorbing states, the pair
contact process (PCP) and the reaction dimer (RD) model.
Both models have a continuous transition from a reactive phase
into an inactive phase with infinitely many absorbing states,
which is shown,
rather surprisingly, again in the DP universality class.

This result might mislead the readers that the number of absorbing states
is not relevant to the universality class of
the absorbing phase transitions.
This shows a sharp contrast
to the case of equilibrium critical phenomena where the
number of ground states plays a crucial role. The symmetry
between absorbing states may be more important than the
number of absorbing states in determining the universality
class. In the PCP, the
infinitely many absorbing states are not equivalent
probabilistically, i.e.~some absorbing states can be reached
more easily than other absorbing states by the PCP dynamics.
So it is important to study a model with multiple
equivalent absorbing states.

A few models have appeared in the literature which
have two equivalent absorbing states.
Those are the model $A$ and $B$ of probabilistic cellular automa (PCA)
introduced by Grassberger {\it et al} \cite{Grass84,Grass89},
nonequilibrium kinetic Ising models with two different
dynamics (NKI) \cite{Meny94,Meny95},
and the interacting monomer-dimer model (IMD) \cite{Park941,Park951}.
The PCA and NKI models are single-component models, while
the IMD model is a multi-component model.
Critical behaviors of these models are different from DP but
seem to belong to the same universality class
(see Table 1). This result
confirms our assertion that the symmetry between absorbing states is
relevant to the universality class of absorbing phase transitions.

Recently, the branching annihilating random walks (BAW) with
offsprings have been studied intensively
\cite{Taka,Sud,Jen931,Jen941}. The BAW models exhibit a
continuous absorbing phase transition from an active steady
state into a single absorbing state. According to the foregoing
DP conjecture,
this phase transition should belong to the DP universality class.
However, recent numerical investigations of the BAW models with even
number of offsprings (BAWe) reveal that the BAWe models
do not belong to the DP class but the same
universality class as in the models with two equivalent
absorbing states \cite{Taka,Jen941}
(Table 1). The BAW models with odd number of
offsprings (BAWo) belong to the conventional DP class \cite{Jen934}.
Dynamics of the BAWe models conserve the number of walkers
modulo 2, while the BAWo models evolve without any conservation
law. So the conservation law seems also relevant to the universality
class of absorbing phase transitions.

  The remaining question is why the models with two equivalent
absorbing states and the BAWe models with the mass (particle number)
conservation of modulo 2 exhibit critical behaviors belonging to the same
universality class.
Grassberger described the PCA models in
terms of kinks and showed that the number of kinks is conserved
modulo 2 in the PCA models \cite{Grass89}.
The NKI model can be also described by kinks (or domain walls)
and its dynamics conserve the number of kinks modulo 2 \cite{Meny94}.
The kink representation of the IMD model is complicated due to
the presence of various
kinks resulting from its multi-component nature.
Each type of kinks has no conservation law but the total number of
kinks is conserved modulo 2 again \cite{Park942}.
So these models bear a close
resemblance to each other, even though there is no exact mapping available
between
them.  From the spirit of universality, one can argue
that differences among the above models are just irrelevant details
which do not affect the universal behavior, e.g.~scaling exponents.

In this paper, we take a careful look at these details and raise
the question of whether models with the same conservation law in its
kink representation always belong to the same universality class.
This question is intimately related to the role of symmetry
between absorbing states in determining the universality class
of absorbing phase transitions.

We consider the
IMD model with infinitely strong repulsions between the
same species in one dimension. A monomer ($A$) cannot adsorb at a
nearest-neighbor site of an already-occupied monomer
(restricted vacancy) but adsorb at a free vacant site with
no adjacent monomer-occupied sites at a rate $k_A$. Similarly,
a dimer ($B_2$) cannot adsorb at a pair of restricted vacancies
($B$ in nearest-neighbor sites) but adsorb at a pair of free
vacancies at a rate $k_B$.
There are no nearest-neighbor restrictions
in adsorbing particles of different species.
Here we will consider only the adsorption-limited reactions. A
nearest neighbor of the adsorbed $A$ and $B$ particles reacts
immediately, forms the $AB$ product, and desorbs the catalytic
surface.
Whenever there is an $A$
adsorption attempt at a vacant site inbetweeen an
adsorbed $A$ and an adsorbed $B$, we allow the $A$ to adsorb
and react immediately with the neighboring $B$, thus forming
an $AB$ product and  desorbing the surface.
The system has no fully saturated phases of
monomers or dimers, but instead two equivalent half-filled absorbing
states. These states, $I$ and $II$, comprise of only the monomers at the
odd- or even-numbered lattice sites. A dimer needs a pair of
adjacent vacancies to adsorb, so a state with
alternating sites occupied by monomers can be identified with an
absorbing state.

In order to study the role of symmetry between
these absorbing states, we introduce a symmetry breaking field
$h$ which
favors one absorbing state over the other.
This can be done
by differentiating the adsorption rate of monomers
at an odd-numbered vacant site and at an even-numbered one.
If a monomer is chosen to adsorb on an even-numbered
free vacant site, the adsorption-attempt is rejected with probability
$h$ ($0\le h\le 1$). The case $h=0$ corresponds to the ordinary
IMD model previously studied in details \cite{Park941,Park951}.
For finite $h$
the monomers tend to adsorb more on an odd-numbered site
than an even-numbered one. Therefore the absorbing state $I$ can be
reached more easily than the other absorbing state $II$ by this dynamics.

In this paper, we set $h=0.5$ for convenience.
Then the system can be characterized by one parameter
$p=k_A/(k_A+k_B)$ of the monomer adsorption-attempt
probability. The dimer adsorption-attempt probability is
given by $q = 1 - p $.
We perform dynamic Monte Carlo simulations for this model.
We start with a lattice occupied by monomers at all
odd-numbered sites except at the central vacant site.
Then the system evolves along the dynamic rules of the model.
After one adsorption attempt on the average per lattice site
(one Monte Carlo step), the time is incremented by one unit.
5000 independent runs
are made up to 5000 time steps
for various values of $p$ near the critical probability $p_c$.
Most runs, however, stop earlier because the system gets into the
absorbing state $I$. We measure the survival probability $P(t)$ (the
probability that the system is still active at time $t$),
the number of dimers $N(t)$ averaged over all
runs, and the mean-square distance of spreading $R^2 (t)$ averaged
over the surviving runs.
At criticality, the values of these quantities scale
algebraically in the long time limit \cite{Grass79}
\begin{eqnarray}
P(t) &\sim& t^{-\delta},\\
N(t) &\sim& t^{\eta},\\
R^2(t) &\sim& t^{z},
\end{eqnarray}
and double-logarithmic plots of these values against time
show straight lines. Off criticality, these plots show
some curvatures.
More precise estimates for the scaling exponents can be obtained
by examining the local slopes of the curves.
The effective
exponent $\delta(t)$ is defined as
\begin{equation}
-\delta(t) = \frac{\log \left[ P(t) / P(t/b) \right]}{\log ~b}
\end{equation}
and similarly for $\eta (t)$ and $z(t)$. In Fig.~1, we plot
the effective exponents against $1/t$ with $b =10$.
Off criticality these plots
show upward or downward curvatures. From Fig.~1, we estimate
$p_c  \simeq  0.414(1)$ which is much lower than in
the ordinary IMD model without the symmetry breaking field \cite{Park951}
($p_c \simeq 0.5325(5)$) as expected.
The scaling exponent is given by the intercept
of the critical curve with the vertical axis. Our estimates for the
dynamic scaling exponents are
\begin{equation}
  \delta =0.160(5) ,~~~ \eta = 0.31(1), ~~~ z = 1.40(20).
\end{equation}
These values are completely different from the values in
the IMD model without the symmetry breaking field
(Table 1), but are in an excellent accord with
those of the DP universality class; $\delta=0.1596(4),
\eta=0.3137(10)$, and $z=1.2660(14)$
(see reference \cite{Jen941}),
although the exponent $z$ shows very slow convergence.
Similar results are obtained as we set the value of
the symmetry breaking field $h=0.1$.

In summary, we have studied the IMD model with infinitely
strong repulsive nearest-neighbor interactions between
the same species in one dimension. This system exhibits a continuous
transition from a reactive phase into an inactive phase
with two equivalent absorbing states. Its critical behavior is
different from the DP universality class but consistent with
that of the PCA, NKI, and BAWe models.
As soon as the symmetry breaking field is introduced,
the two absorbing states of the IMD model become
probabilistically not equivalent and
this model exhibits a critical behavior
in the DP universality class. It implies that the
symmetry between absorbing states plays a crucial role
in determining the universality class.
Our numerical results indicate that the system evolves mostly into
one favored absorbing state, not bothered by the other
unfavored absorbing state. So the critical behavior
may be governed by a fixed point with a single absorbing
state, which is associated with the DP universality class.
This may explain why the PCP model with infinitely many
absorbing states also belongs to the DP universality class.
For more numerical evidences, we are currently performing
static Monte Carlo simulations for this symmetry-broken model
and also studying
the role of the symmetry breaking field in the
PCA and NKI models.
It is interesting to note that
the total number of kinks in the IMD model is still conserved modulo 2
in the presence of the symmetry breaking field.
So it is incorrect to say that all models with mass conservation
of modulo 2 should belong to the same universality class other than
the DP class.

This work is supported in part by NON DIRECTED
RESEARCH FUND, Korea Research Foundation and by the BSRI, Ministry of
Education (Grant No.~95-2409).

\newpage
\begin{center}
{\Large {\bf Table Caption}}
\vspace{5mm}
\end{center}
\begin{description}
\item[{\bf Table 1 :}]
Numerical estimates of scaling exponents. $\eta_d$ ($\eta_i$) is the
exponent for the defect (interface) dynamics \cite{Park951}. These
estimates are taken from references
\cite{Grass89,Meny94,Park941,Park951,Jen941}. Numbers in
parentheses represent the errors in the last digits.
\end{description}
\begin{center}
\begin{tabular}{ccccc}
\hline\hline
exponent & PCA & NKI & IMD & BAWe \\ \hline
$\delta$ & 0.283(16) & 0.27(2) & 0.29(2)& 0.285(2) \\
$\eta_d$ &  --- & --- & 0.00(2) & 0.000(1) \\
$\eta_i$ & 0.272(12)& 0.30(2) & 0.285(20) & 0.282(4)\\
$z$  & 1.11(2) & 1.14(2) & 1.14(2) &1.141(2) \\
$\nu_\parallel$ & 3.3(2) & 2.9(6) & 3.17(5) & 3.25(10) \\
$\beta$ & 0.94(6) & 0.80(8) & 0.88(3) & 0.92(3) \\ \hline\hline
\end{tabular}
\end{center}

\newpage
{\center\Large Figure Captions}
\begin{description}
   \item[Fig.\  1  :]  Plots of the effective exponents against
    $1/t$. Threee curves from top to bottom
    in each panel correspond to $p=0.413,$ 0.414, 0.415.
    Thick lines are critical lines ($p=0.414$).

\end{description}

\end{document}